\documentclass[12pt]{article}
\usepackage{graphicx}
\usepackage{float}

\begin{document}

\begin{flushright}

IMSc/2020/02/01

\end{flushright} 

\vspace{2mm}

\vspace{2ex}

\begin{center}

{\large \bf Non singular, bouncing M theory universe}

\vspace{8ex}

{\large  Saroj Prasad Chhatoi and S. Kalyana Rama}

\vspace{3ex}

Institute of Mathematical Sciences, HBNI, C. I. T. Campus, 

\vspace{1ex}

Tharamani, CHENNAI 600 113, India. 

\vspace{2ex}

email: sarojpc, krama at imsc dot res dot in \\

\end{center}

\vspace{6ex}

\centerline{ABSTRACT}

\begin{quote} 

We propose a set of equations as a simple model for non singular
evolutions of a $10 + 1$ dimensional M theory universe. Our
model uses ideas from Loop Quantum Cosmology and offers a
solution to the important problem of singularity resolutions. We
solve the equations numerically and find that an M theory
universe in this model evolves non singularly and with a bounce
: going back in time, its density reaches a maximum and
decreases thereafter whereas its physical size reaches a non
vanishing minimum and increases thereafter. Taking the
constituents of the universe to be the most entropic ones (which
are four sets of intersecting M branes) leads to an effectively
$3 + 1$ dimensional spacetime as the M theory universe expands,
both in the infinite past and future.

\end{quote}














\newpage






The $9 + 1 $ dimensional superstring theory, equivalently the
$10 + 1 $ dimensional M theory, is a candidate for a quantum
theory of gravity and is also expected to describe the matter
contents and the quantum evolution of our universe. Consider our
universe. It is $3 + 1$ dimensional and, at densities and
temperatures small compared to Planckian ones, its cosmological
evolution is well described by general relativity equations for
a homogeneous isotropic universe. However, general relativity
equations lead to a big bang singularity in the past where the
densities and temperatures exceed Planckian values and diverge
to infinity.

Such singularities are expected to be resolved upon quantising
gravity, thus within string/M theory. See \cite{bowick} --
\cite{m08} for several string/M theoretic ideas for resolving
the big bang singularities. To mention an example : String
theory has a T duality symmetry under which the radius $R$ of a
compact direction is transformed to its inverse, namely $R \to
l_s^2 \; R^{- 1}$ where $l_s$ is the string length. For an
universe whose spatial directions may all be taken as circles,
one then expects that the sizes of all the circles will be
bounded below by $l_s$ and, hence, that the curvature of the
universe will be bounded above by $\simeq l_s^{- 2} \;$. Such a
limiting curvature may then resolve the big bang singularities.

Enormous progress has been made in string/M theory towards, for
example, understanding the entropy and the Hawking radiation of
extremal and near extremal black holes. However, no comparable
progress has been made towards understanding the big bang
singularities.  The stringy mechanisms resolving the big bang
singularities are still not understood in full detail. Nor is
there any simple model leading to non singular evolution of a
string/M theory universe. Also, a successful model should lead
to an effectively $3 + 1$ dimensional spacetime as the string/M
theory universe expands. See \cite{bv}, \cite{tv}, \cite{dks} --
\cite{k10} for several ideas for obtaining a $3 + 1$ dimensional
universe in string/M theory. 

An alternative candidate for a quantum theory of gravity is the
$3 + 1$ dimensional Loop Quantum Gravity (LQG) constructed using
Ashtekar variables \cite{ashtekar} -- \cite{bkrv}. It leads to
Loop Quantum Cosmology (LQC) upon restricting to homogeneous
variables of cosmology and quantising them. The resulting
quantum evolutions of the $3 + 1$ dimensional universe have been
extensively studied and found to resolve the big bang
singularities \cite{b} -- \cite{lb}. It has also been found that
these non singular quantum evolutions are well described by a
set of effective equations which reduce to general relativity
equations in the `classical limit'.

Recently, one of us have empirically generalised the effective
LQC equations to $d + 1$ dimensional universes and to include
arbitrary functions; and then studied analytically their salient
features \cite{k16} -- \cite{k19}. In this letter, we will
propose these generalised effective equations as a simple model
for the evolution of a $10 + 1$ dimensional M theory universe.
The model will have one arbitrary function $f(x) \;$. General
relativity equations follow for $f(x) = x \;$. We will take
$f(x) = sin \; x \;$ and solve the effective equations
numerically. The resulting evolution of an M theory universe
will be non singular and have a bounce. Namely, as one goes back
in time, the density of the universe will reach a maximum and
then start decreasing thereafter. Correspondingly, the physical
size of the universe will reach a non vanishing minimum, bounce
back, and then start increasing thereafter. Our model thus uses
ideas from LQC, applies them in an M theory context, and offers
a solution to the important problem of singularity resolutions. 

The line element $d s$ for an M theory universe is taken to be
given by
\begin{equation}\label{ds}
d s^2 = - \; d t^2 + \sum_i e^{2 \lambda^i} \; (d x^i)^2
\end{equation}
where $i = 1, 2, \cdots, 10$ and the scale factors
$e^{\lambda^i}$ are functions of $t$ only. Let $\rho$ and $p_i$
be the total density of the constituents of the universe and
their total pressures in the $i^{th}$ direction. Define the
quantities $G_{i j}, \; G^{i j}, \; \Lambda \;$, and $r^i \;$ by
\begin{eqnarray}
G_{i j} \; = \; 1 - \delta_{i j} & , & G^{i j} \; = \;
\frac{1}{9} - \delta^{i j} \; \; , \nonumber \\
& & \nonumber \\
\Lambda \; = \; \sum_i \lambda^i & , & 
r^i \; = \; \sum_j G^{i j} \; (\rho - p_j) \; \; ; \label{ri}
\end{eqnarray}
let $f(x)$ be the function which characterises the model and
which is required to $\to x$ as $x \to 0$ but is arbitrary
otherwise; let $m^i \; , \; i = 1, 2, \cdots, 10$ be a new set
of variables, to be related to the time derivatives of the scale
factors using $f(x) \;$; and, define the functions $f^i, \;
g_i$, and $X_i$ by
\begin{equation}\label{fgx} 
f^i = f(m^i) \; \; , \; \; \;
g_i = \frac{d \; f^i} {d m^i} \; \; , \; \; \;
X_i = g_i \sum_j G_{i j} f^j \; \; . 
\end{equation}
Then we propose that the equations governing the evolution of
the scale factors $e^{\lambda^i}$ be given by \cite{inspired} 
\begin{eqnarray}
l_{qm} \; \lambda^i_t & = & \sum_j G^{i j} X_j \; \; ,
\label{e1} \\
& & \nonumber \\
\sum_{i j} G_{i j} f^i f^j & = & 2 \; l_{qm}^2 \; 
\kappa^2 \; \rho \; \; , \label{e2} 
\end{eqnarray}
\begin{equation}\label{e3}
(m^i)_t \; + \; \sum_j \frac {(m^i - m^j) \; X_j}
{9 \; l_{qm}} \; = \; l_{qm} \; \kappa^2 \;
\left( r^i - \frac {2 \rho} {9} \right) 
\end{equation}
or equivalently
\begin{equation}\label{e4}
(m^i)_t \; + \; \Lambda_t \; m^i \; = \; l_{qm} \; \kappa^2 \;
r^i \; + \; \sum_{j k} \frac {G_{j k} \; (m^j g_j - f^j) \; f^k}
{9 \; l_{qm}} \; \; , 
\end{equation}
and the standard conservation equation
\begin{equation}\label{rhot} 
\rho_t \; + \; \sum_i (\rho + p_i) \; \lambda^i_t \; = \; 0
\; \; . 
\end{equation}
Equation (\ref{rhot}) follows from equations (\ref{ri}) --
(\ref{e3}); equivalently, one may take equation (\ref{e2}) to
follow from other equations. In the above equations, $\kappa^2 =
8 \pi G_{11} \;$, the $t-$subscripts denote the time
derivatives, and $l_{qm} = {\cal O} (1) \; \kappa^{ \frac {2}
{9}} $ is a length parameter which, in LQC, would characterise
the quantum of area. The $10 + 1$ dimensional general relativity
equations \cite{k10} follow in the `classical limit' where $f(x)
= x \;$, hence $g_i = 1$ and $l_{qm} \lambda^i_t = f^i = m^i
\;$; more generally, any linear function $f(x) = b x + c \;$
where $b$ and $c$ are constants gives $l_{qm} \lambda^i_t = b
f^i$ and the general relativity equations with $\kappa^2$
replaced by $b^2 \kappa^2 \;$.  Substituting for $\rho$ and
$p_i$ the densities and the pressures of the constituents of an
M theory universe will then give its evolution in a model
specified by a function $f(x) \;$.

Consider the constituents of an M theory universe. It is natural
to assume that they must be the most entropic ones \cite{k206}.
As explained lucidly in \cite{m06, m08}, such constituents are
$N$ sets of brane configurations which intersect according to
the Bogomol'nyi -- Prasad -- Sommerfeld (BPS) rules, with
highest possible $N \;$ \cite{bps} -- \cite{g}. According to the
BPS rules, two stacks of M5 branes intersect along three common
spatial directions; two stacks of M2 branes intersect along zero
common spatial directions; a stack of M2 branes intersect a
stack of M5 branes along one common spatial direction; and each
stack of branes is smeared uniformly along the other brane
directions. There can be a wave along the common intersection
direction \cite{bps, b2ps, b3ps}. High entropies of these
configurations arise due to the phenomenon of fractionation of
branes \cite{m06, m08}. Requiring atleast three spatial
directions to be not wrapped by any intersecting branes, so that
an M theory universe may resemble ours, then restricts $N$ to be
$\le 4 \;$ \cite{gkt}. Thus $N = 4$ for the most entropic
configurations which, with no loss of generality, we may take to
be given by four stacks of intersecting M branes which wrap the
seven directions, labelled $1, 2, \cdots, 7 \;$: namely, two
stacks each of $M 2$ and $M 5$ branes wrap respectively the
directions $1 2$, $\; 3 4$, $\; 1 3 5 6 7$, and $2 4 5 6 7 \;$.

The densities $\rho_I$ and the pressures $p_{i I}$ of the
$I^{th}$ stack of branes, $I = 1, 2, \cdots, N \;$, in such BPS
configurations are mutually noninteracting and seperately
conserved. Thus 
\begin{equation}\label{rhoIt} 
\rho \; = \; \sum_I \rho_I \; \; , \; \; \;
p_i \; = \; \sum_I p_{i I} \; \; , \; \; \;
(\rho_I)_t + \sum_i (\rho_I + p_{i I}) \; \lambda^i_t \; = \; 0
\; \; .
\end{equation}
To proceed further, equations of state are needed which
determine the pressures $p_{i I}$ in terms of $\rho_I \;$. For
black brane configurations, they follow from the M theory
action. For cosmology, they were derived from first principles
in \cite{m06, m08} under certain assumptions. One may also show
that the U--duality symmetries of M theory require that the
density $\rho_{(I)}$ of the $I^{th}$ stack and its pressures
$p_{\parallel (I)}$ and $p_{\perp (I)}$ along the parallel and
transverse directions must be related as follows \cite{k207,
k08, k10} :
\begin{equation}\label{pp}
p_{\parallel (I)} = - \rho_{(I)} + 2 \; p_{\perp (I)} \; \; .
\end{equation}
Specifying $p_{\perp (I)}$ as a function of $\rho_{(I)}$ will
determine the equations of state for $p_{\parallel (I)}$ and
thereby for all the pressures $p_{i (I)} \;$. The U--duality
symmetries further require this function to be the same for all
$I \;$. Hence, specifying a single function $p_\perp (\rho)$
determines all $p_{i I}$ in terms of $\rho_I$ where $i = 1, 2,
\cdots, 10$ and $I = 1, 2, \cdots, N \;$. The result derived in
\cite{m06, m08} follows as a special case where $p_\perp (\rho)
= 0 \;$. More generally, we assume that $p_\perp (\rho) = (1 -
u) \; \rho$ where $u$ is a constant. For the $N = 4$ case, with
the four stacks of branes denoted by $I = 2, \; 2', \; 5, \; 5'
\;$, the pressures $p_{i I}$ are then given by \cite{k207, k08,
k10}
\begin{eqnarray} 
\{ (\rho - p_i)_{(2)} \} \; :
& (2, \; 2, \; 1, \; 1, \; 1, \; 1, \; 1, \; 1, \; 1, \; 1) \; 
u \; \rho_{(2)} & , \nonumber \\
& & \nonumber \\
\{ (\rho - p_i)_{(2')} \} \; :
& (1, \; 1, \; 2, \; 2, \; 1, \; 1, \; 1, \; 1, \; 1, \; 1) \; 
u \; \rho_{(2')} & , \nonumber \\
& & \nonumber \\
\{ (\rho - p_i)_{(5)} \} \; :
& (2, \; 1, \; 2, \; 1, \; 2, \; 2, \; 2, \; 1, \; 1, \; 1) \; 
u \; \rho_{(5)} & , \nonumber \\
& & \nonumber \\
\{ (\rho - p_i)_{(5')} \} \; :
& (1, \; 2, \; 1, \; 2, \; 2, \; 2, \; 2, \; 1, \; 1, \; 1) \;
u \; \rho_{(5')} & ;  \label{pi22'55'} 
\end{eqnarray}
and the corresponding $r^i_I = \sum_j G^{i j} \; (\rho_I - p_{j
I})$ by
\begin{eqnarray}
\{ r^i_{(2)} \} \; : & (- 2, \; - 2, \; 1, \; 1, \; 1, \; 1, \;
1, \; 1, \; 1, \; 1) \;
\frac {u \; \rho_{(2)}} {3} & , \nonumber \\
& & \nonumber \\
\{ r^i_{(2')} \} \; : & (1, \; 1, \; - 2, \; - 2, \; 1, \; 1, \;
1, \; 1, \; 1, \; 1) \;
\frac {u \; \rho_{(2')}} {3} & , \nonumber \\
& & \nonumber \\
\{ r^i_{(5)} \} \; : & (- 1, \; 2, \; - 1, \; 2, \; - 1, \; - 1,
\; - 1, \; 2, \; 2, \; 2) \;
\frac {u \; \rho_{(5)}} {3} & , \nonumber \\
& & \nonumber \\
\{ r^i_{(5')} \} \; : & (2, \; - 1, \; 2, \; - 1, \; - 1, \; -
1, \; - 1, \; 2, \; 2, \; 2) \;
\frac {u \; \rho_{(5')}} {3} & . \label{ri22'55'}
\end{eqnarray}
Thus, the simple model we propose for the evolution of an M
theory universe is given by equations (\ref{e1}) --
(\ref{rhoIt}) with $f(x) = sin \; x$ and by the equations of
state (\ref{pi22'55'}) or equivalently (\ref{ri22'55'}).

We obtain the evolution of an M theory universe in our model by
numerically solving equations (\ref{e1}) -- (\ref{rhoIt}) for
$\lambda^i (t), \; m^i (t)$, and $\rho_I (t) \;$. We consider
the $N = 4$ case, which is the most entropic one, as well as the
$N < 4$ cases by taking one or more $\rho_I$ to vanish. In our
numerical studies, we set $l_{qm} = \kappa^2 = 1$ with no loss
of generality by measuring the time in units of $l_{qm} \;$ and
the densities and the pressures in units of $l_{qm}^{- 2} \;
\kappa^{- 2} \;$; set $\lambda^i = 0$ for all $i$ at an initial
time $t_0 \;$; and, for the sake of definiteness, take $u =
\frac {2} {3} \;$ which corresponds to $p_\perp = \frac {\rho}
{3} \;$. We then obtain the numerical solutions for $\lambda^i
(t), \; m^i (t) \;$, and $\rho_I (t)$ for various sets of
initial values $m^i_0 = m^i (t_0) $ and $\rho_{I 0} = \rho_{I}
(t_0) \;$; some of the $\rho_{I 0} = 0$ for the $N < 4$ cases.
The results we deduce from our numerical solutions are listed
below and they show clearly that the evolutions are non singular
and have bounces.

\begin{itemize}

\item

The densities $\rho_I (t)$ and the total density $\rho (t)$ have
finite maxima. In the limit $t \to \pm \infty \;$, all the non
vanishing $\rho_I$ become equal to each other and $\to 0 \;$.
See Figure {\bf 1}.

\item

The total volume factor $e^{\Lambda (t)}$ has a non vanishing
minimum and $\to \infty \;$ as $t \to \pm \; \infty \;$. See the
plot of $\Lambda (t)$ in Figure {\bf 2}.

\item

For all $i \; , \; m^i (t)$ remain finite, $\; \to 0$ as $t \to
\infty \;$, and $\to \pi \;$ as $t \to - \; \infty \;$. See
Figure {\bf 3}. In the limit $t \to \pm \infty$ then the
function $f(x) = sin \; x$ is linear and, hence, the evolutions
are as in general relativity.

\item

For all $i \; , \; \lambda^i_t (t)$ remain finite. See Figure
{\bf 4}. In the limit $t \to \pm \; \infty \;$, the scale
factors $e^{\lambda^i} \sim \vert t \vert^{\alpha^i}$ and $t \;
\lambda^i_t \sim \alpha^i$ where $\alpha^i$ are constants.

\vspace{2ex}

The exponents $\alpha^i$ can be calculated from general
relativity equations when the non vanishing $\rho_I$ are equal
to each other. Depending on which of the $\rho_I$ are non
vanishing, the exponents $\alpha^i$ for different $i$ may be
negative, vanishing, or positive. Then, in the limit $t \to \pm
\; \infty \;$, the corresponding $\lambda^i$ may $\to - \infty,
\; c^i_\pm \;$, or $\infty $ where $c^i_\pm$ are constants; the
corresponding scale factor $e^{\lambda^i}$ may or may not have a
bounce. See the plots of $\lambda^i (t)$ and $t \; \lambda^i_t
(t)$ in Figures {\bf 5 -- 7}.

\end{itemize}

We illustrate these features in figures {\bf 1} -- {\bf 7} in a
few select cases. In these figures, $u = \frac {2} {3} \;$, the
initial time $t_0 = e^{- 10} \;$, and the initial values are :
$\lambda^i (t_0) = 0 \;$; $\left\{ m^i_0 \right\} = (-24, \; 05,
\; 52, \; 91, \; -78, \; 22, \; 32, \; 42, \; 67, \; -29) *
(0.01)$ in figures {\bf 1 -- 5} and $= (-24, \; 20, \; 52, \;
91, \; -78, \; 10, \; 32, \; 42, \; 67, \; -29) * (0.01) \;$ in
figures {\bf 6} and {\bf 7}; and, $\left\{ \rho_{I 0} \right\}
\propto ( 4.75, \; 21.8, \; 183, \; 373) \;$ in figures {\bf 1
-- 5}, $\propto ( 4.75, \; 0.0, \; 0.0, \; 373) \;$ in figure
{\bf 6}, and $\propto ( 1.0, \; 0.0, \; 0.0, \; 0.0) \;$ in
figure {\bf 7}. The proportionality constants in $\left\{
\rho_{I 0} \right\}$ are fixed by requiring that equation
(\ref{e2}) be satisfied at $t_0 \;$. In the figures, we have not
labelled the curves with their $i$ or $I$ indices since such
labellings are not illuminating for our purposes here. 

\begin{figure}[H]\label{figrho}
\centering
\includegraphics[scale=0.95]{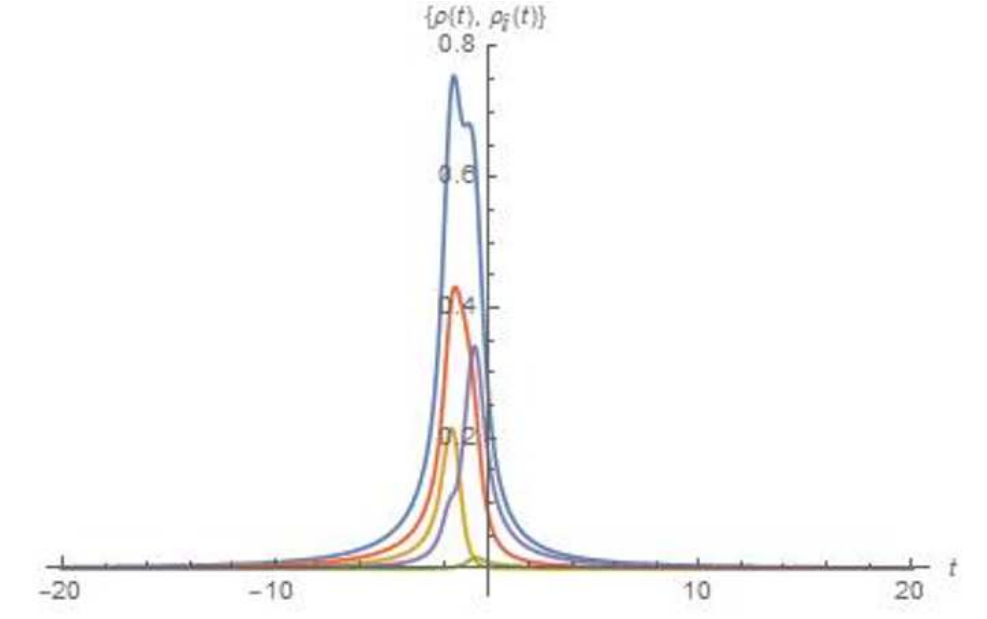}
\caption{
Plots of total $\rho(t)$ and $\rho_I (t)$ for all $I$ showing
their finite maxima. All four $\rho_I$ are non vanishing.
}
\end{figure}

\begin{figure}[H]\label{figl}
\centering
\includegraphics[scale=0.95]{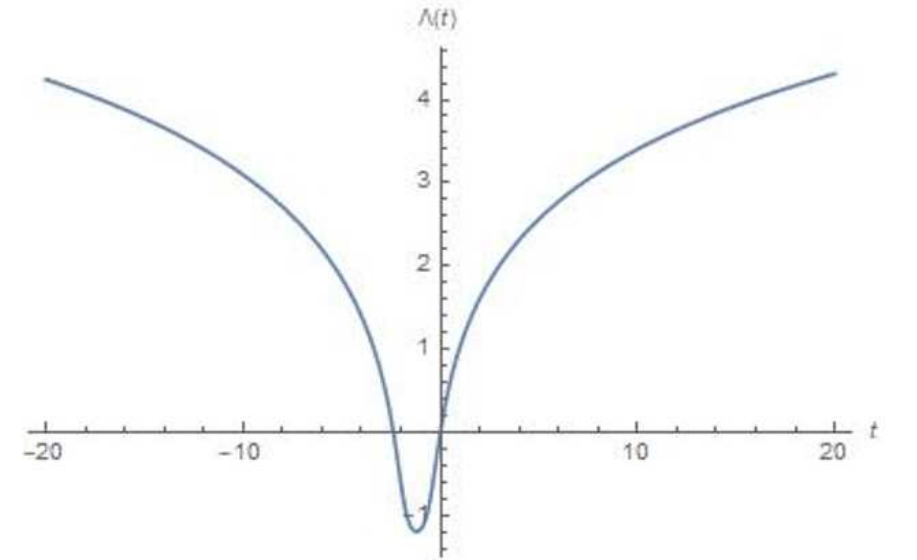}
\caption{
Plot of $\Lambda (t)$ showing its finite minimum and bounce.
All four $\rho_I$ are non vanishing.
}
\end{figure}

\begin{figure}[H]\label{figmi}
\centering
\includegraphics[scale=0.95]{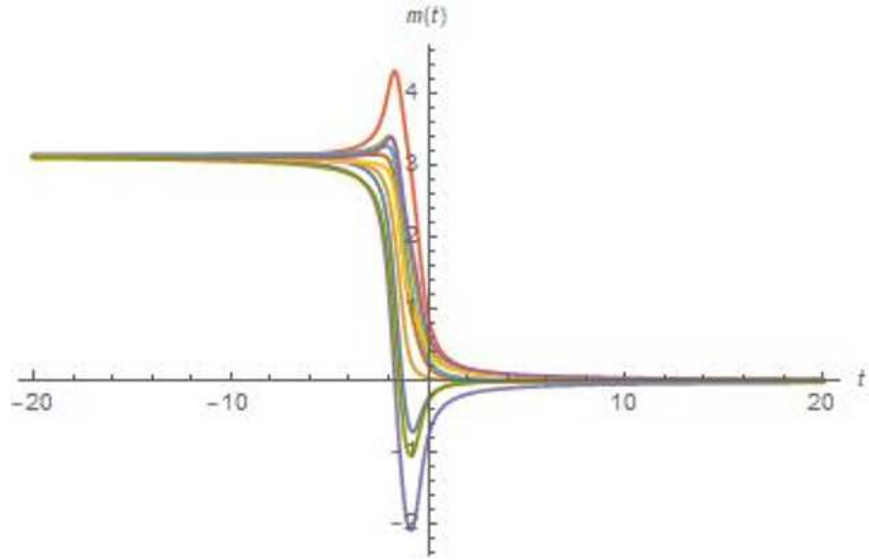}
\caption{
Plots of $m^i (t)$ for all $i \;$ showing their finiteness and
their approach to $0$ as $t \to \infty$ and to $\pi \;$ as $t
\to - \infty \;$. All four $\rho_I$ are non vanishing.
}
\end{figure}

\begin{figure}[H]\label{figlit}
\centering
\includegraphics[scale=0.95]{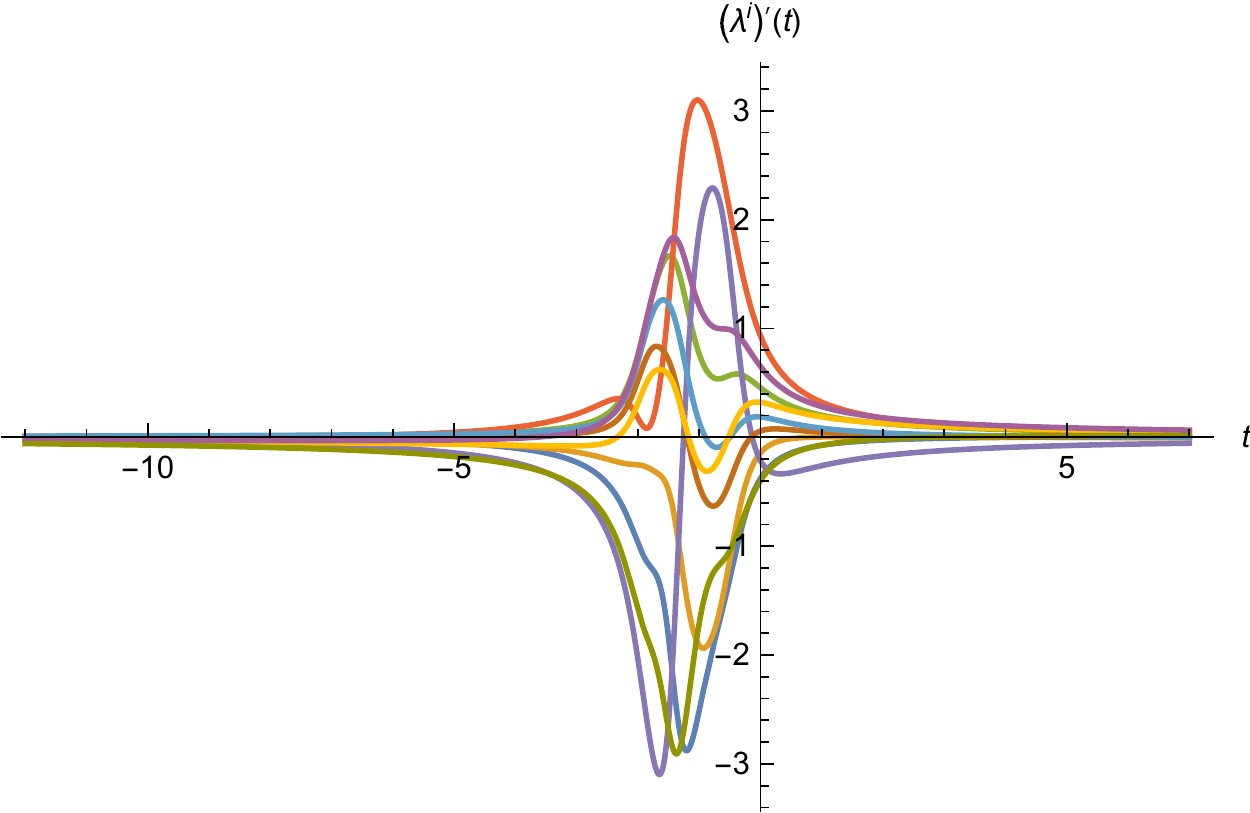}
\caption{
Plots of $\lambda^i_t (t)$ for all $i$ showing their finiteness.
All four $\rho_I$ are non vanishing.
}
\end{figure}

The plots of total $\rho, \; \rho_I, \; \Lambda, \; m^i \;$, and
$\lambda^i_t \;$, shown above in figures {\bf 1 -- 4}, are
qualitatively similar in all the cases we studied. The plots of
$\lambda^i$ and $t \lambda^i_t$ can be different and are shown
in figures {\bf 5 -- 7} when all four $\rho_I$ are non
vanishing, only $\rho_{(2)}$ and $\rho_{(5')}$ are non
vanishing, and when only $\rho_{(2)}$ is non vanishing. The
abscissae in these figures are $ln \; t$ for the figures on the
right hand side and $ln \; (- t)$ for the figures on the left
hand side. In the limit $t \to \pm \infty \;$, the non vanishing
$\rho_I$ become equal to each other, the scale factors
$e^{\lambda^i} \to \vert t \vert^{\alpha^i} \;$, hence $t
\lambda^i_t \to \alpha^i \;$, and the evolutions are as in
general relativity. The exponents $\left\{ \alpha^i \right\}$
can then be calculated analytically. They are given, after a
straightforward calculation, by $(0, 0, 0, 0, 0, 0, 0, 1, 1, 1)
* \frac {1} {2}$ in figure {\bf 5}, $(0, - 1, 1, 0, 0, 0, 0, 1,
1, 1) * \frac {3} {7}$ in figure {\bf 6}, and by $(- 2, - 2, 1,
1, 1, 1, 1, 1, 1, 1) * \frac {1} {4}$ in figure {\bf 7}.

\begin{figure}[H]\label{figli4}
\centering
\includegraphics[scale=.64]{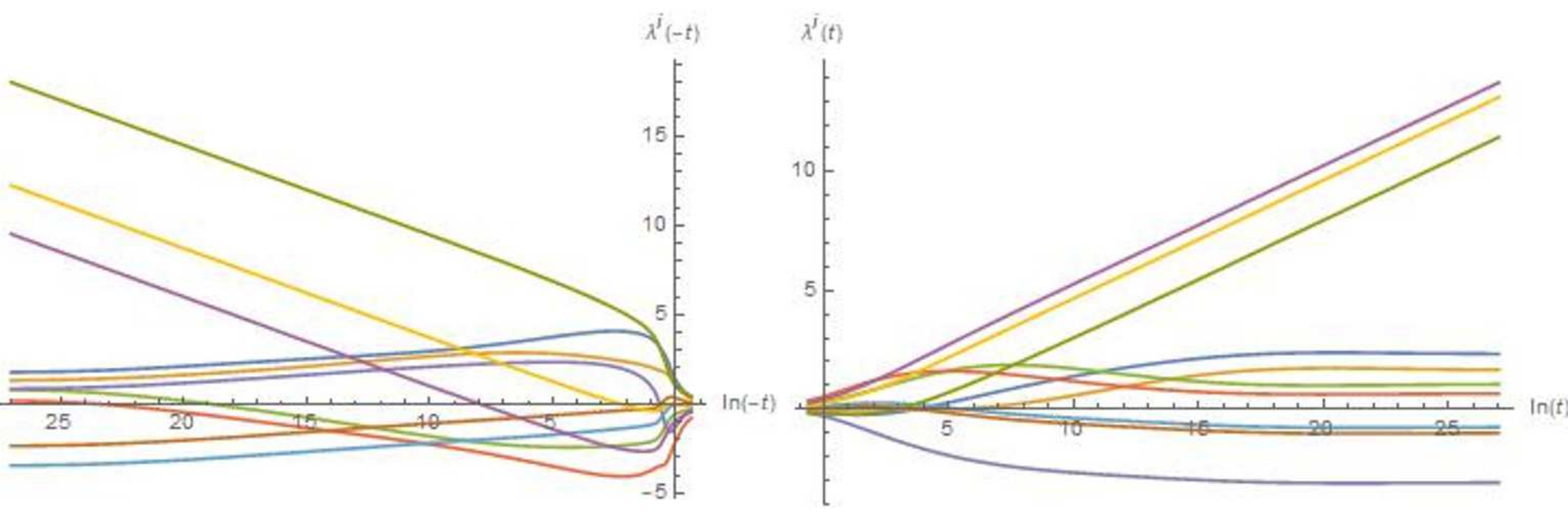}
\end{figure}

\begin{figure}[H]\label{figlit4}
\centering
\includegraphics[scale=.74]{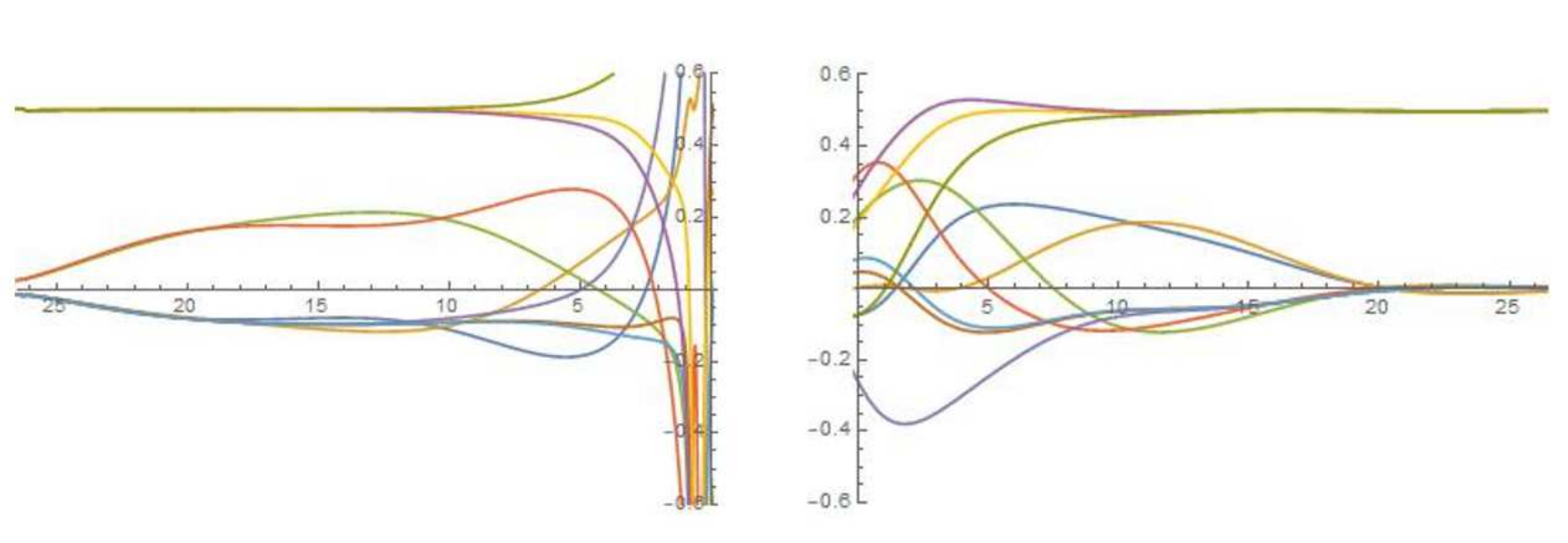}
\caption{
Plots of $\lambda^i (t)$ (upper panel) showing their approach to
$\infty$ or $c^i_\pm \;$, and of $t \lambda^i_t (t)$ (lower
panel) showing their approach to $\alpha^i \;$, as $t \to \pm
\infty \;$. Analytically, $\left\{ \alpha^i \right\} = (0, 0, 0,
0, 0, 0, 0, 1, 1, 1) * \frac {1} {2} \;$.  All four $\rho_I$ are
non vanishing. }
\end{figure}

\newpage

\begin{figure}[H]\label{figli2}
\centering
\includegraphics[scale=.65]{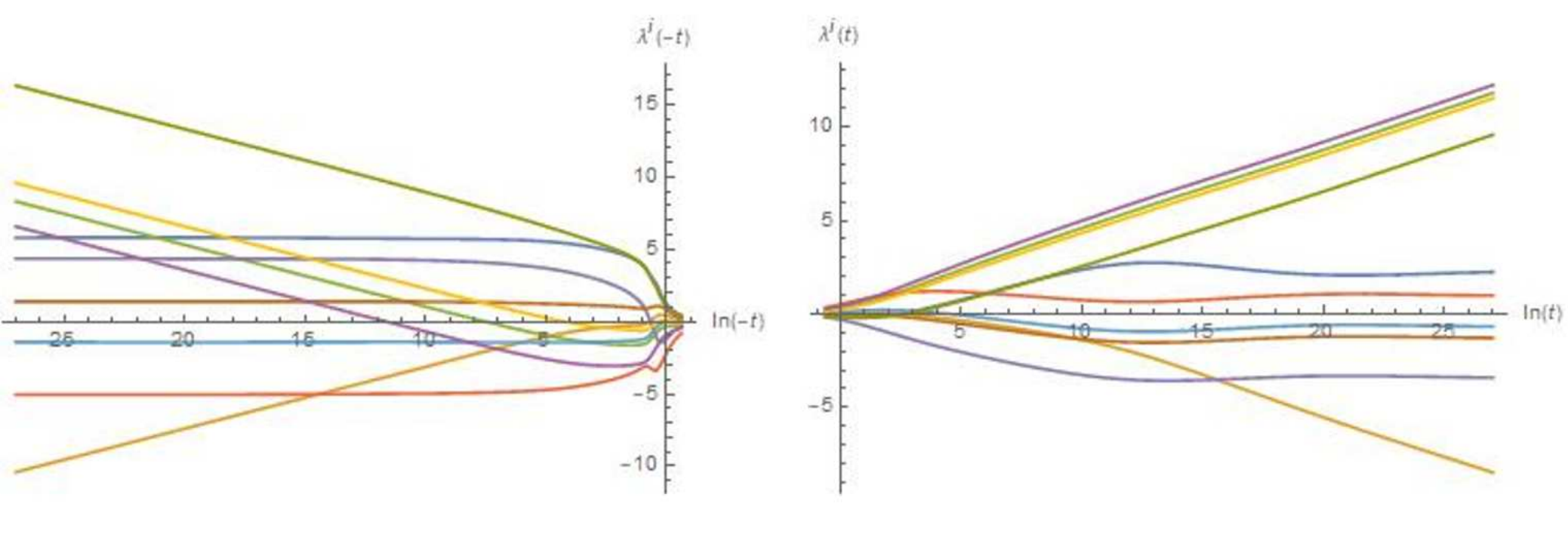}
\end{figure}

\begin{figure}[H]\label{figlit2}
\centering
\includegraphics[scale=.74]{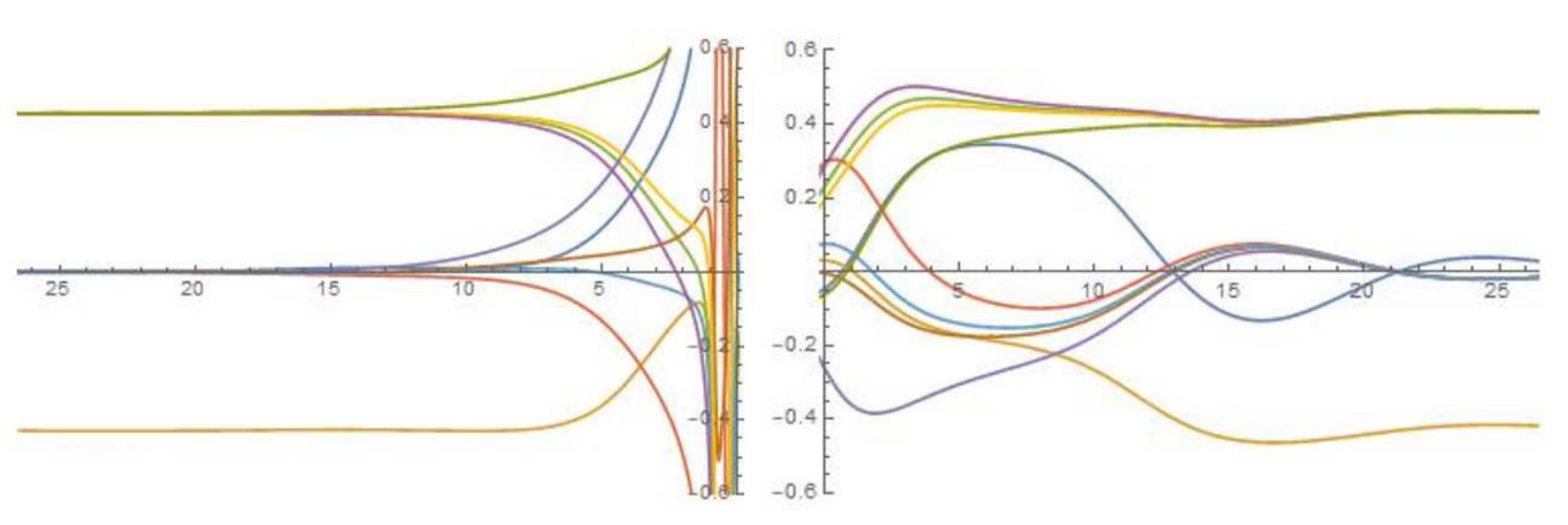}
\caption{
Plots of $\lambda^i (t)$ (upper panel) showing their approach to
$\pm \infty$ or $c^i_\pm \;$, and of $t \lambda^i_t (t)$ (lower
panel) showing their approach to $\alpha^i \;$, as $t \to \pm
\infty \;$. Analytically, $\left\{ \alpha^i \right\} = (0, - 1,
1, 0, 0, 0, 0, 1, 1, 1) * \frac {3} {7} \;$. Only $\rho_{(2)}$
and $\rho_{(5')}$ are non vanishing.  }
\end{figure}

\newpage

\begin{figure}[H]\label{figli1}
\centering
\includegraphics[scale=.66]{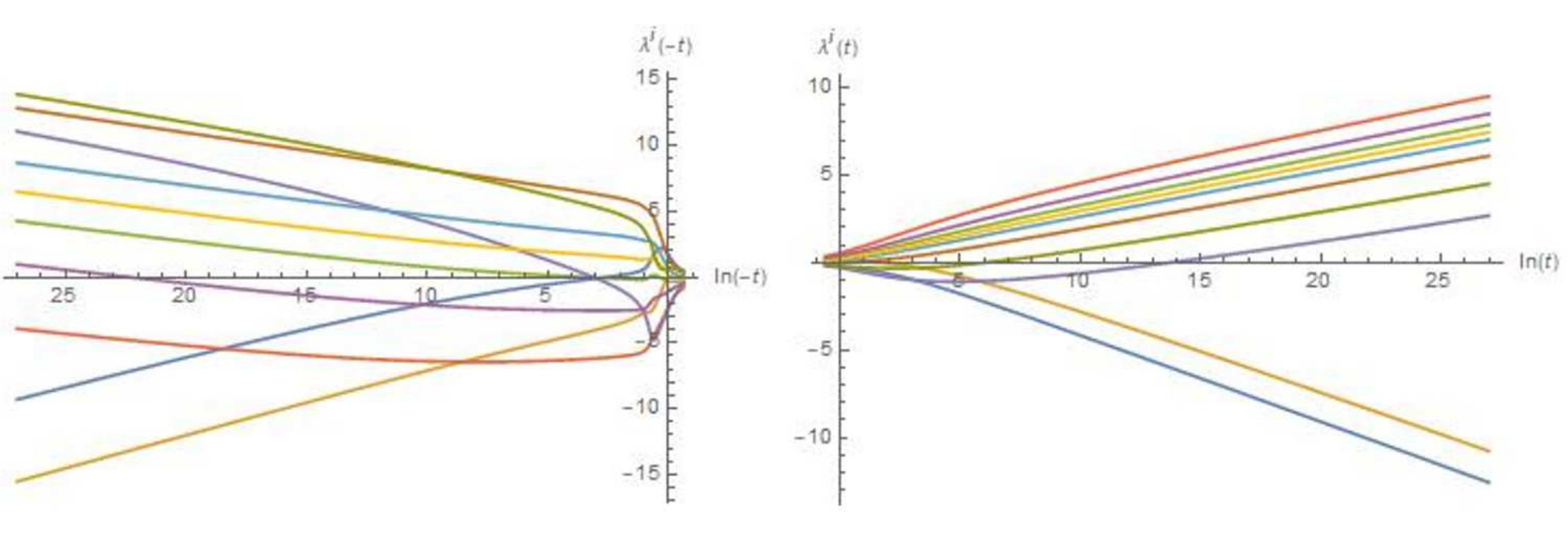}
\end{figure}

\begin{figure}[H]\label{figlit1}
\centering
\includegraphics[scale=.74]{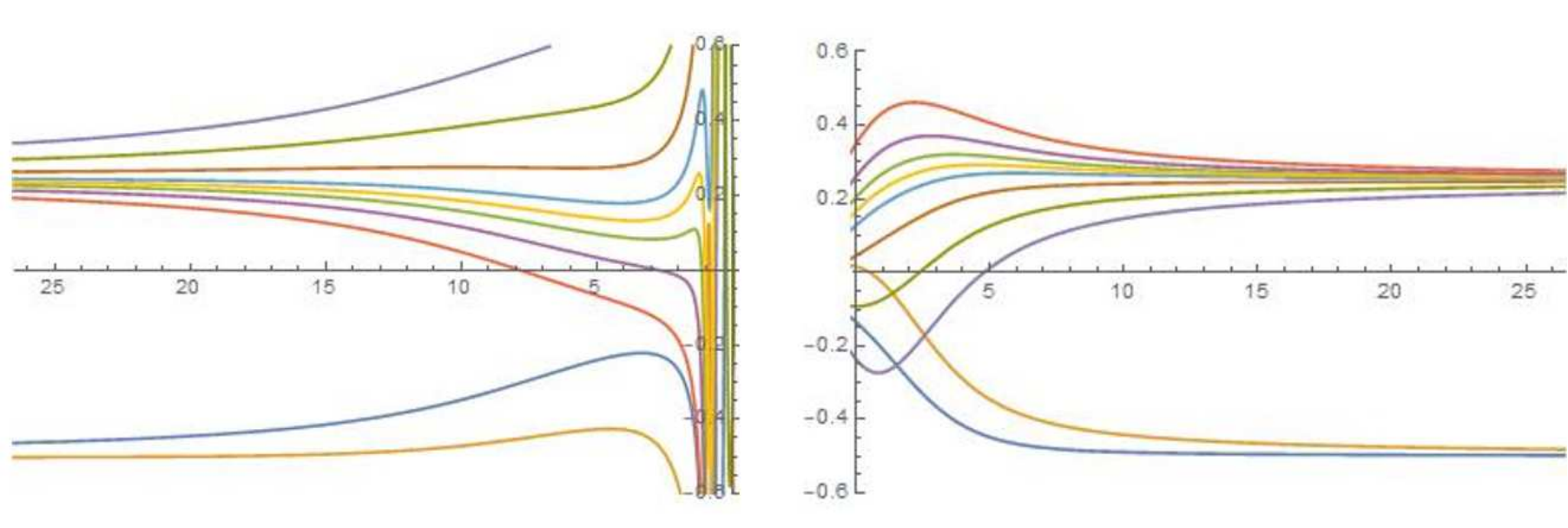}
\caption{
Plots of $\lambda^i (t)$ (upper panel) showing their approach to
$\pm \infty \;$, and of $t \lambda^i_t (t)$ (lower panel)
showing their approach to $\alpha^i \;$, as $t \to \pm \infty
\;$. Analytically, $\left\{ \alpha^i \right\} = (- 2, - 2, 1, 1,
1, 1, 1, 1, 1, 1) * \frac {1} {4} \;$. Only $\rho_{(2)}$ is non
vanishing. }
\end{figure}

The behaviour of $\lambda^i$ in figures {\bf 5 -- 7} in the
limit $t \to \pm \infty$ can also be explained physically as
follows \cite{k10} : In this general relativity limit, $m^i
\propto \lambda^i_t$ and, hence, the terms $r^i_I$ on the right
hand side of equation (\ref{e3}) may be thought of as a force
due to the $I^{th}$ stack of branes on $\lambda^i \;$. It
follows from equations of state (\ref{ri22'55'}) that, when
$\rho_I$ are equal to each other, directions parallel to M2 or
M5 branes experience a contracting force of strength $2$ or $1$
in some units, and directions transverse to them experience an
expanding force of strength $1$ or $2 \;$. Therefore,
$\lambda^i$ for the directions with net contracting or vanishing
or expanding force will $\to - \infty \;$, or $\to c^i_\pm$ or
$\to \infty $ in the limit $t \to \pm \infty \;$.

Thus, if only $\rho_{(2)}$ is non vanishing then $\lambda^i \to
- \infty$ at the rate of 2 in some units for $i = 1, \; 2$ and
$\to \infty$ at the rate of 1 for the remaining $i \;$, thus
leading to an effectively $8 + 1$ dimensional spacetime in the
limit $t \to \pm \infty \;$. If only $\rho_{(2)}$ and
$\rho_{(5')}$ are non vanishing then $\lambda^i \to - \infty$ at
the rate of 3 for $i = 2 \;$, $\; \to c^i_\pm$ for $i = 1, \; 4,
\; 5, \; 6, \; 7 \;$, and $\to \infty$ at the rate of 3 for $i =
3, \; 8, \; 9, \; 10 \;$, thus leading to an effectively $4 + 1$
dimensional spacetime in the limit $t \to \pm \infty \;$. If all
the four $\rho_I$ are non vanishing, as in the most entropic
case, then $\lambda^i \to c^i_\pm$ for $i = 1, \; \cdots, 7 \;$
and $\to \infty$ at the rate of 6 for $i = 8, \; 9, \; 10 \;$,
thus leading to an effectively $3 + 1$ dimensional spacetime in
the limit $t \to \pm \infty \;$ \cite{k08, k10}. These features
can be seen clearly in Figures {\bf 5 -- 7}.


In summary, we have presented a simple model leading to non
singular evolutions of a $10 + 1$ dimensional M theory universe.
Our model uses ideas from LQC and offers a solution to the
important problem of singularity resolutions. Also, modelling
the M theory constituents in the most entropic case as in
\cite{m06, m08, k08, k10} leads to an effectively $3 + 1$
dimensional spacetime as the M theory universe expands.

There are several avenues for further studies. One may study
numerically, or analytically where possible, the evolution of
higher dimensional universes within our model but for different
functions, for example $f(x) = tanh \; x \;$. It seems possible
\cite{zhang, work} to derive the present model for which $f(x) =
sin \; x \;$ using the higher dimensional formulation of LQG
given in \cite{th1} -- \cite{th3}. But it is not clear which
other functions may be allowed in such an approach, see
\cite{h}. One may study higher derivative actions which may lead
to the effective equations of the present model, perhaps
following the approach of \cite{bo}. Such actions may then be
used for, among other things, studying inhomegeneous
perturbations and their evolutions in non singular
universes. One may also analyse systematically whether or not
string/M theoretic higher derivative actions lead to non
singular evolutions of an universe; and, if yes, study their
similarities and differences with the non singular evolutions
seen in this letter.



\begin{thebibliography}{999}

\bibitem{bowick}




M.~J.~Bowick and L.~C.~R.~Wijewardhana, 
{\em Superstrings at High Temperature,}
Phys.\ Rev.\ Lett.\  {\bf 54}, 2485 (1985), \\
doi:10.1103/PhysRevLett.54.2485. 

\bibitem{b2owick}
M.~J.~Bowick and L.~C.~R.~Wijewardhana, 
{\em Superstring Gravity and the Early Universe,} 
Gen.\ Rel.\ Grav.\  {\bf 18}, 59 (1986), \\
doi:10.1007/BF00843749. 

\bibitem{bv}
R.~H.~Brandenberger and C.~Vafa,
{\em Superstrings in the Early Universe,}
Nucl.\ Phys.\ B {\bf 316}, 391 (1989), 
doi:10.1016/0550-3213(89)90037-0. 
  
\bibitem{tv}
A.~A.~Tseytlin and C.~Vafa,
{\em Elements of string cosmology,} \\
Nucl.\ Phys.\ B {\bf 372}, 443 (1992),  \\
doi:10.1016/0550-3213(92)90327-8,  
[hep-th/9109048].

\bibitem{gv}
M.~Gasperini and G.~Veneziano,
{\em Pre - big bang in string cosmology},
Astropart.\ Phys.\  {\bf 1}, 317 (1993), \\
doi:10.1016/0927-6505(93)90017-8, 
[hep-th/9211021].

\bibitem{k97}
S. Kalyana Rama,
{\em Can string theory avoid cosmological singularities?},
Phys.\ Lett.\ B {\bf 408}, 91 (1997), \\
doi:10.1016/S0370-2693(97)00795-8, 
[hep-th/9701154].

\bibitem{mr}
M.~Maggiore and A.~Riotto,
{\em D-branes and cosmology,} \\
Nucl.\ Phys.\ B {\bf 548}, 427 (1999), \\
doi:10.1016/S0550-3213(99)00104-2, 
[hep-th/9811089].

\bibitem{bfm}
T.~Banks, W.~Fischler and L.~Motl, 
{\em Dualities versus singularities,}
JHEP {\bf 01}, 019 (1999), \\
doi:10.1088/1126-6708/1999/01/019, 
[hep-th/9811194].

\bibitem{gv2}
M.~Gasperini and G.~Veneziano,
{\em The Pre - big bang scenario in string cosmology},
Phys.\ Rept.\  {\bf 373}, 1 (2003), \\
doi:10.1016/S0370-1573(02)00389-7, 
[hep-th/0207130].

\bibitem{nayeri}
A.~Nayeri, R.~H.~Brandenberger and C.~Vafa, 
{\em Producing a scale-invariant spectrum of perturbations in a
Hagedorn phase of string cosmology,}
Phys.\ Rev.\ Lett.\  {\bf 97}, 021302 (2006), \\
doi:10.1103/PhysRevLett.97.021302, 
[hep-th/0511140].

\bibitem{k06}
S.~Kalyana Rama,
{\em A Stringy correspondence principle in cosmology,}
Phys.\ Lett.\ B {\bf 638}, 100 (2006),  \\
doi:10.1016/j.physletb.2006.05.047,  
[hep-th/0603216]. 

\bibitem{m06}
B.~D.~Chowdhury and S.~D.~Mathur,
{\em Fractional Brane State in the Early Universe,} 
Class.\ Quant.\ Grav.\  {\bf 24}, 2689 (2007), \\
doi:10.1088/0264-9381/24/10/014, 
[hep-th/0611330].

\bibitem{m08}
S.~D.~Mathur,
{\em What is the state of the Early Universe?}, \\
J.\ Phys.\ Conf.\ Ser.\  {\bf 140}, 012009 (2008), \\
doi:10.1088/1742-6596/140/1/012009, 
[arXiv:0803.3727 [hep-th]].






\bibitem{dks}
R.~Durrer, M.~Kunz and M.~Sakellariadou,
{\em Why do we live in 3+1 dimensions?},
Phys.\ Lett.\ B {\bf 614}, 125 (2005), \\
doi:10.1016/j.physletb.2005.04.023
[hep-th/0501163].

\bibitem{kr}
A.~Karch and L.~Randall,
{\em Relaxing to three dimensions},
Phys.\ Rev.\ Lett.\  {\bf 95}, 161601 (2005), \\
doi:10.1103/PhysRevLett.95.161601, 
[hep-th/0506053].

\bibitem{k206}
S.~Kalyana Rama,
{\em A Principle to Determine the Number (3 + 1) of Large
Spacetime Dimensions,}
Phys.\ Lett.\ B {\bf 645}, 365 (2007), \\
doi:10.1016/j.physletb.2006.11.077, 
[hep-th/0610071].

\bibitem{k207}
S.~Kalyana Rama,
{\em Consequences of U dualities for Intersecting Branes in the
Universe,}
Phys.\ Lett.\ B {\bf 656}, 226 (2007), \\
doi:10.1016/j.physletb.2007.09.069, 
[arXiv:0707.1421 [hep-th]].

\bibitem{k08}
S. Bhowmick, S. Digal and S. Kalyana Rama, \\
{\em Stabilisation of Seven (Toroidal) Directions and Expansion
of the remaining Three in an M theoretic Early Universe Model,}
\\
Phys.\ Rev.\ D {\bf 79}, 101901 (2009), \\
doi:10.1103/PhysRevD.79.101901, 
[arXiv:0810.4049 [hep-th]].

\bibitem{k10}
S. Bhowmick and S. Kalyana Rama, \\
{\em 10 + 1 to 3 + 1 in an Early Universe with mutually BPS
Intersecting Branes}, 
Phys.\ Rev.\ D {\bf 82}, 083526 (2010), \\
doi:10.1103/PhysRevD.82.083526, 
[arXiv:1007.0205 [hep-th]].






\bibitem{ashtekar}
A.~Ashtekar,
{\em New Variables for Classical and Quantum Gravity,} \\
Phys.\ Rev.\ Lett.\  {\bf 57}, 2244 (1986),  
doi:10.1103/PhysRevLett.57.2244.

\bibitem{a2shtekar}
A.~Ashtekar,
{\em New Hamiltonian Formulation of General Relativity,} \\
Phys.\ Rev.\ D {\bf 36}, 1587 (1987), 
doi:10.1103/PhysRevD.36.1587. 

\bibitem{lqg}
A.~Ashtekar and J.~Lewandowski,
{\em Background independent quantum gravity: A Status report},
Class.\ Quant.\ Grav.\  {\bf 21}, R53 (2004), \\
doi:10.1088/0264-9381/21/15/R01, 
[gr-qc/0404018].

\bibitem{bka}
A. Ashtekar,
{\em Lectures on non-perturbative canonical gravity}, \\
Notes prepared in collaboration with R. S. Tate, \\
World Scientific, Singapore (1991).

\bibitem{bkr}
C. Rovelli,
{\em Quantum Gravity}, \\
Cambridge University Press, Cambridge (2004).

\bibitem{bkt}
T. Thiemann, \\
{\em Introduction to modern canonical quantum general
relativity}, \\
Cambridge University Press, Cambridge (2005). 

\bibitem{bkrv}
C. Rovelli and F. Vidotto,
{\em Covariant loop quantum gravity}, \\
Cambridge University Press, Cambridge (2014).






\bibitem{b}
M.~Bojowald,
{\em Absence of singularity in loop quantum cosmology,} \\
Phys.\ Rev.\ Lett.\  {\bf 86}, 5227 (2001),  \\
doi:10.1103/PhysRevLett.86.5227,  
[gr-qc/0102069].

\bibitem{b01}
M.~Bojowald,
{\em The Inverse scale factor in isotropic quantum geometry,} \\
Phys.\ Rev.\ D {\bf 64}, 084018 (2001), \\
doi:10.1103/PhysRevD.64.084018, 
[gr-qc/0105067]. 

\bibitem{b02}
M.~Bojowald,
{\em Isotropic loop quantum cosmology,} \\ 
Class.\ Quant.\ Grav.\  {\bf 19}, 2717 (2002),  \\
doi:10.1088/0264-9381/19/10/313, 
[gr-qc/0202077]. 

\bibitem{b03}
M.~Bojowald,
{\em Homogeneous loop quantum cosmology,} \\
Class.\ Quant.\ Grav.\  {\bf 20}, 2595 (2003), \\
doi:10.1088/0264-9381/20/13/310, 
[gr-qc/0303073].

\bibitem{abl}
A.~Ashtekar, M.~Bojowald and J.~Lewandowski, 
{\em Mathematical structure of loop quantum cosmology,} 
Adv.\ Theor.\ Math.\ Phys.\  {\bf 7}, 233 (2003), \\
doi:10.4310/ATMP.2003.v7.n2.a2, 
[gr-qc/0304074].

\bibitem{aps}
A.~Ashtekar, T.~Pawlowski and P.~Singh, \\
{\em Quantum nature of the big bang,} 
Phys.\ Rev.\ Lett.\  {\bf 96}, 141301 (2006), \\
doi:10.1103/PhysRevLett.96.141301, 
[gr-qc/0602086].

\bibitem{aps2}
A.~Ashtekar, T.~Pawlowski and P.~Singh, 
{\em Quantum Nature of the Big Bang: Improved dynamics,} 
Phys.\ Rev.\ D {\bf 74}, 084003 (2006), \\
doi:10.1103/PhysRevD.74.084003, 
[gr-qc/0607039].

\bibitem{aw}
A.~Ashtekar and E.~Wilson-Ewing, 
{\em Loop quantum cosmology of Bianchi I models,} 
Phys.\ Rev.\ D {\bf 79}, 083535 (2009), \\
doi:10.1103/PhysRevD.79.083535, 
[arXiv:0903.3397 [gr-qc]].

\bibitem{h}
R.~C.~Helling,
{\em Higher curvature counter terms cause the bounce in loop
cosmology},
arXiv:0912.3011 [gr-qc].

\bibitem{bo}
C.~Barragan and G.~J.~Olmo,
{\em Isotropic and Anisotropic Bouncing Cosmologies in Palatini
Gravity},
Phys.\ Rev.\ D {\bf 82}, 084015 (2010), \\
doi:10.1103/PhysRevD.82.084015
[arXiv:1005.4136 [gr-qc]].

\bibitem{status}
A.~Ashtekar and P.~Singh, 
{\em Loop Quantum Cosmology: A Status Report,} \\
Class.\ Quant.\ Grav.\  {\bf 28}, 213001 (2011), \\
doi:10.1088/0264-9381/28/21/213001,  
[arXiv:1108.0893 [gr-qc]].

\bibitem{lb}
L.~Linsefors and A.~Barrau,
{\em Modified Friedmann equation and survey of solutions in
effective Bianchi-I loop quantum cosmology}, \\
Class.\ Quant.\ Grav.\  {\bf 31}, 015018 (2014), \\
doi:10.1088/0264-9381/31/1/015018
[arXiv:1305.4516 [gr-qc]].






\bibitem{k16}
S. Kalyana Rama,
{\em A Class of LQC--inspired Models for Homogeneous,
Anisotropic Cosmology in Higher Dimensional Early Universe,} \\
Gen.\ Rel.\ Grav.\  {\bf 48}, 155 (2016), \\
doi:10.1007/s10714-016-2150-2, 
[arXiv:1608.03231 [gr-qc]].

\bibitem{k17}
S. Kalyana Rama,
{\em Variety of $(d + 1)$ dimensional Cosmological Evolutions
with and without bounce in a class of LQC -- inspired Models,} \\
Gen.\ Rel.\ Grav.\  {\bf 49}, 113 (2017), \\
doi:10.1007/s10714-017-2277-9, 
[arXiv:1706.08220 [gr-qc]].

\bibitem{k18}
S. Kalyana Rama, 
{\em Isotropic LQC and LQC-inspired Models with a massless
scalar field as Generalised Brans-Dicke theories}, \\
Gen.\ Rel.\ Grav.\  {\bf 50}, 56 (2018), \\
doi:10.1007/s10714-018-2378-0, 
[arXiv:1802.06349 [gr-qc]].

\bibitem{k19}
S. Kalyana Rama, 
{\em Non singular M theory Universe in Loop Quantum Cosmology --
inspired Models},
Gen.\ Rel.\ Grav.\  {\bf 51}, 75 (2019), \\
doi:10.1007/s10714-019-2556-8, 
[arXiv:1903.09770 [hep-th]].

\bibitem{inspired}

Equations (\ref{e1}) -- (\ref{e3}) were obtained in a different
and more general form in \cite{k16} by generalising the
effective LQC equations in a simple, empirical, but natural
way. Their salient features were studied analytically in
\cite{k17} -- \cite{k19}. They give the effective LQC equations
for $d = 3$ and $f(x) = sin \; x \;$. The $d = 3$ case with
other $f(x) \;$ has been envisaged in \cite{h}. Using the higher
dimensional formulation of LQG given in \cite{th1} --
\cite{th3}, the $d > 3$ case with $f(x) = sin \; x \;$ has been
derived in \cite{zhang} for an isotropic universe; a similar
derivation for an anisotropic universe also appears to be
possible \cite{work}.






\bibitem{th1}
N.~Bodendorfer, T.~Thiemann and A.~Thurn, 
{\em New Variables for Classical and Quantum Gravity in all
Dimensions I. Hamiltonian Analysis,} \\
Class.\ Quant.\ Grav.\  {\bf 30}, 045001 (2013), \\
doi:10.1088/0264-9381/30/4/045001, 
[arXiv:1105.3703 [gr-qc]].

\bibitem{th2}
N.~Bodendorfer, T.~Thiemann and A.~Thurn,
{\em New Variables for Classical and Quantum Gravity in all
Dimensions II. Lagrangian Analysis,} \\
Class.\ Quant.\ Grav.\  {\bf 30}, 045002 (2013), \\
doi:10.1088/0264-9381/30/4/045002, 
[arXiv:1105.3704 [gr-qc]].

\bibitem{th3}
N.~Bodendorfer, T.~Thiemann and A.~Thurn,
{\em New Variables for Classical and Quantum Gravity in all
Dimensions III. Quantum Theory,} \\
Class.\ Quant.\ Grav.\  {\bf 30}, 045003 (2013), \\
doi:10.1088/0264-9381/30/4/045003, 
[arXiv:1105.3705 [gr-qc]].

\bibitem{zhang}
X.~Zhang,
{\em Higher dimensional Loop Quantum Cosmology,} \\
Eur.\ Phys.\ J.\ C {\bf 76}, 395 (2016), \\
doi:10.1140/epjc/s10052-016-4249-8, 
[arXiv:1506.05597 [gr-qc]].

\bibitem{work}
S. Kalyana Rama and Arnab Priya Saha, {\em Unpublished notes.} 






\bibitem{bps}
A.~A.~Tseytlin,
{\em Harmonic superpositions of M-branes,} \\
Nucl.\ Phys.\ B {\bf 475}, 149 (1996), \\
doi:10.1016/0550-3213(96)00328-8, 
[hep-th/9604035].

\bibitem{b2ps}
M.~Cvetic and A.~A.~Tseytlin, 
{\em Nonextreme black holes from nonextreme intersecting
M-branes,}
Nucl.\ Phys.\ B {\bf 478}, 181 (1996), \\
doi:10.1016/0550-3213(96)00411-7, 
[hep-th/9606033].

\bibitem{b3ps}
A.~A.~Tseytlin,
{\em `No force' condition and BPS combinations of p-branes in
eleven-dimensions and ten-dimensions,} \\
Nucl.\ Phys.\ B {\bf 487}, 141 (1997), \\
doi:10.1016/S0550-3213(96)00692-X, 
[hep-th/9609212].






\bibitem{gkt}
J.~P.~Gauntlett, D.~A.~Kastor and J.~H.~Traschen,
{\em Overlapping branes in M theory,} 
Nucl.\ Phys.\ B {\bf 478}, 544 (1996), \\
doi:10.1016/0550-3213(96)00423-3, 
[hep-th/9604179].

\bibitem{betal}
E.~Bergshoeff, M.~de Roo, E.~Eyras, B.~Janssen and J.~P.~van der
Schaar,
{\em Multiple intersections of D-branes and M-branes,} \\
Nucl.\ Phys.\ B {\bf 494}, 119 (1997), \\
doi:10.1016/S0550-3213(97)00151-X, 
[hep-th/9612095].

\bibitem{g}
J.~P.~Gauntlett,
{\em Intersecting branes,}
In *Seoul/Sokcho 1997, Dualities in gauge and string theories*
146-193, \\
doi:10.1142/9789814447287\underline{ }0004, 
[hep-th/9705011].







\end{thebibliography}
\end{document}